\documentclass[aps,pre,amsmath,amsfonts,amssymb,floatfix,superscriptaddress,nofootinbib]{revtex4}
\pdfoutput=1
\usepackage{mathrsfs} \usepackage{graphicx,color} 
\usepackage{bbm} \usepackage{multirow}
\let\a=\alpha \let\b=\beta  
   
\let\l=\lambda    
\let\s=\sigma \let\t=\tau  
   
\let\D=\Delta   
   
 \let\r=\rho

\def\de{\mathrm d}
\newcommand{\E}{{\mathbb E}}
\newcommand{\uS}{\underline{S}}

\def\to{\rightarrow}

\newcommand{\beq}{\begin{equation}} \newcommand{\eeq}{\end{equation}}

\begin{document}

\title{
Quasi equilibrium construction for the long time limit of glassy dynamics
} 

\author{Silvio Franz}
\affiliation{LPTMS, CNRS, Univ. Paris Sud, Université Paris-Saclay, 91405 Orsay, France
}

\author{Giorgio Parisi}
\affiliation{Dipartimento di Fisica,
INFN, Sezione di Roma I, and NANOTEC -- CNR,
Sapienza Universit\'a di Roma,
P.le A. Moro 2, I-00185 Roma, Italy
}

\author{Federico Ricci-Tersenghi}
\affiliation{Dipartimento di Fisica,
INFN, Sezione di Roma I, and NANOTEC -- CNR,
Sapienza Universit\'a di Roma,
P.le A. Moro 2, I-00185 Roma, Italy
}

\author{Pierfrancesco Urbani}
\affiliation{Institut de physique théorique, Universit\'e Paris Saclay, CEA, CNRS, F-91191, Gif-sur-Yvette}

\begin{abstract}
  In this paper we review a recent proposal to understand
the long time limit of glassy dynamics in terms of
  an appropriate Markov Chain.
  \cite{FP13}.  The advantages of the resulting construction are many. The first
  one is that it gives a quasi equilibrium description on how glassy
  systems explore the phase space in the slow relaxation part of their
  dynamics. The second one is that it gives an alternative way to
  obtain dynamical equations starting from a dynamical rule that is
  static in spirit. This provides a way to overcome the difficulties
  encountered in the short time part of the dynamics where current
  conservation must be enforced. We study this approach in detail in a prototypical
  mean field disordered spin system, namely the $p$-spin spherical
  model, showing how we can obtain the well known equations that describes its dynamics.  
  Then we apply the same approach to structural glasses. We first derive a set of dynamical Ornstein-Zernike
  equations which are very general in nature. Finally we consider two possible closure schemes for them, namely the Hypernetted Chain approximation
  of liquid theory and a closure of the BBGKY hierarchy that has been recently introduced by G. Szamel.
  From both approaches we finally find a set of dynamical Mode-Coupling like equations that are supposed to describe the system in the long time/slow dynamics regime.
\end{abstract}

\maketitle

\section{Introduction}

Glass is an out of equilibrium state of matter. Let by themselves, the
microscopic configurations and macroscopic properties of glasses
slowly change in a process known as physical aging, where the
free-energy evolves towards lower and lower values. This process is
very slow and is characterized by a separation of time scales, where
fast degrees of freedom appear to be in thermal equilibrium in the
background of the slow degrees of freedom. The evolution of the latter
is sometime described as an effective random walk in configuration
space, depicted as a rough free-energy landscape, where the system
wanders from one metastable state to another. The rules by which the
metastable states are selected in the dynamical process determine the
observed properties of the systems. A leading role in the
comprehension of slow off-equilibrium dynamics of glassy systems is
played by mean field theory (see \cite{leticia} for a review), which
allows to describe asymptotic aging regimes through scaling laws and
effective temperatures associated to the violations of the fluctuation
dissipation theorem \cite{CK93,CKP97}.  The analysis performed in
\cite{FV00,Fr03} and in \cite{BKLS00, FM94, Fr94, FMPP98, FMPP99}
related these effective temperature to quasi-equilibrium selection of
metastable states during glassy dynamics. This notion of
quasi-equilibrium exploration can be formalized through the
introduction of a suitable Markov Chain where fast times are
effectively coarse grained, and it is assumed the set of states
available at any give time, which are the ones at a specific distance
from the present state, are selected by a Boltzmann law \cite{FP13}.  This chain
construction reproduces the results of long time relaxational dynamics
within mean-field theory and 
it gives the right long time dynamical equations whenever the system has a finite configurational entropy.
We believe that it captures the principles of exploration of
configuration space in glasses also in realistic systems where metastable states are sufficiently long-lived.  
Moreover, since the short timescales are completely coarse grained, the method automatically produces time reparametrization invariant equations.
Thanks to the Boltzmann prescription, the chain share many formal
features with equilibrium systems. This observation straightforwardly
suggests simple ways to treat long time dynamics of structural glassy systems 
taking advantages of standard approximation schemes originally devised to
study equilibrium \cite{FPU15}. Moreover, since short times are coarse-grained,
constraints such as energy and mass conservation that complicate short
time analysis are not relevant here.  In this contribution we review the
main results of our approach and present some new results and
derivations that were just hinted in previous
publications. The paper is organized as follows: in the
first section we introduce the basic dynamical construction and we discuss the
response properties of the system and its equilibrium measure. In the
second section we discuss spin glass mean field theory, and we give a
new derivation of the dynamical equations through a probabilistic
analysis. This avoids the use of replicas which were used in a previous
publication \cite{FP13}. Then we present the results of the direct
integration of our equations that were never published before. In the
third section we review the applications to realistic liquid
models. We present the derivation of a dynamical version of the Ornstein-Zernike equation. 
We complement this equation with two closure
schemes, that give both a final dynamical equation that is of the same kind of standard Mode-Couling Theory (MCT).
Within this scheme we are able to predict the properties of the dynamics both in the equilibrium regime and in the aging one.

\section{The Boltzmann Pseudodynamics  construction} 
\label{general-considerations} In this section we
review the Boltzmann Pseudodynamics construction recently introduced
in \cite{FP13} and we show some generalities about correlation and
response functions that can be computed using this formalism.  Let us consider a system described by a set of internal degrees of freedom that we call
$S_i$ and that will be addressed as spin variables (our notation is very close to the one encountered in spin
systems but can also be used to treat particles in a liquid where the
internal degrees of freedom are the position of the particles). In the
following we will define a dynamical rule to evolve such system so
that we will indicate with $S_i(t)$ the configuration of the system at
time $t$.  The Boltzmann Pseudodynamics (BPD) is a discrete time dynamics
defined from the following dynamical rule: given the configuration of
the spins at time
$t$, the configuration at time $t+1$ occur with a probability that is
given by 
\beq\label{BPD}
\begin{split}
M(\underline S (t+1)|\underline S(t))&=\frac{1}{Z[\beta_{t+1}; \underline S(t)]}\mathrm{e}^{-\beta_{t+1}H[\underline S(t+1)]}\delta\left(\widetilde C(t,t+1)-q\left(\underline S(t),\underline S({t+1})\right)\right)\\
Z[\beta_{t+1}; \underline S(t)]&=\sum_{\underline S(t+1)}\mathrm{e}^{-\beta_{t+1}H[\underline S(t+1)]}\delta\left(\widetilde C(t,t+1)-q\left(\underline S(t),\underline S({t+1})	\right)\right)
\end{split}
\eeq 
The function $q(\s,\t)$ is an overlap function that measures the
similarity between the configurations $\s$ and $\t$. For spin system it
is given simply by $q(\s,\t)=\sum_i\s_i\t_i/N$ where $\s_i$ and $\t_i$
are the values of the spin in the two configurations. For particle systems instead 
one can choose between many different definitions. A simple possibility is
to use the same function as for spin systems through a lattice gas coding of the liquid
configurations e.g. by dividing the volume in small cells and defining
binary variables that code for occupation numbers of the cells\footnote{Another popular choice in systems of particles is to use
\begin{eqnarray}
  \label{eq:10}
  q(X,Y)=\frac 1 N \sum_{i,j} w(|x_i-y_j|)
\end{eqnarray}
where $X=\{ x_1,...,x_n\}$ and $Y=\{
y_1,...,y_N\}$,  are two configurations of $N$ particles and where $w$ is some positive short ranged function,
e.g. $w(r)=\exp(-r^2/\sigma^2)$.}.
The probability of a trajectory given an initial configuration $\underline S(0)$ at time $t=0$ is
given by 
\begin{eqnarray}
  \label{eq:2}
P[\underline S(t), \underline S(t-1),\ldots ,\underline 
S(1) | \underline S(0)]=\hat P[\underline S(0)] \prod_{k=1}^{t-1}
M(\underline S(k+1)|\underline S(k))   
\end{eqnarray}
where $\hat P$ is a given measure over the inital conditions. Note that in the above
dynamics we fix the set of variables $\{\widetilde C(t+1,t)\}$ and
$\{\b_t\}$ from the outside. In glassy dynamics the temperatures are
naturally fixed to the value of the thermal bath, while $\widetilde C(t+1,t)$
should be chosen in a self-consistent way in order to achieve the
right separation between fast and slow time scales. While in certain
applications it can be interesting to consider a time dependence of
the temperature, from now on in this paper we will consider the case
of $\beta$ constant in time.  We will show in the following that this pseudodynamics provides a coarse grained description of
real time dynamics in which fast processes are seen as instantaneous.  For
finite values of the choosen values $\widetilde C(t+1,t)$, at 
each time step then the system chooses a configuration at a
macroscopic distance from the previous one. This configuration is
chosen as an equilbrium one at the prescribed distance.  In this sense
the ``fast'' time scales that in real dynamics are needed to
equilibrate within a metastable state are coarse-grained. The main assumption is that all configurations satisfying the constraint $q=\widetilde{C}$ are equally reachable by the fast relaxation processes. This assumption is  fine if $\widetilde{C}$ is properly chosen: e.g.\ a value close to the typical overlap $q_{EA}$.

A computer implementation of (\ref{BPD}) would require for each step of the chain
a Montecarlo simulation in which the fast time scale is reinserted to
achieve the prescribed sampling. For that reason we call it
pseudodynamics.  We will be interested to the long chain limit in
which the relevant time scales are much larger than the unit of the
elementary time step and the system moves to even larger distances than the
ones reached in a single step. 

\subsection{Response functions} 
A fundamental characterization of a glassy dynamics is provided by the
linear response functions. Their relation with fluctuations during
aging shows, both in mean-field theory \cite{CK93,CK94} and in simulations
\cite{P97}
the emergence of effective temperatures ruling the exchanges of energy
between slow degrees of freedom.  It has been proposed in \cite{BBM96}
that non trivial response and effective temperatures in
non-equilibrium dynamics are possible if trajectories (clones) starting in the
same point and subject to different thermal noise  get
separated during their evolution. One can define the clone correlation
function of a simple observable as for example the magnetization $m(\uS(t))$ as follows
\[
Q_s(t,u)\equiv\E_{\uS(s)}\left[\E(m(t)|\uS(s)) \E(m(u)|\uS(s)) \right]\quad \text{for } t,u>s.
\]
The internal expectations denoted by $\E(\cdot)$ are the averages over independent thermal trajectories that start from
the same initial configuration $\uS(s)$ at time $s$, while $\E_{\uS(s)}$ is the average over these initial configurations. 
 In \cite{BBM96} it was put forward
the conjecture that non trivial effective temperatures, are only
possible for systems such that for large time separation $Q_s(t,u)$
tends to the same minimal value as the correlation $C(t,u)$
itself. This is different from systems undergoing domain coarseing in
phase separation where $Q$ remains much larger then $C$ and a non
trivial response in the aging regime is absent.  Unfortunately despite
empirical evidence, a formal relation between response and clone
correlation function was lacking. In the dynamics we just introduced
this relation emerges naturally for pseudo-times $u=s+1$ and $t>u$ corresponding to real times $t\gg u$ and $s,u$ such that $C(u,s)$ is in the beta relaxation regime.  Consider the dynamics (\ref{BPD}) in
a time dependent field $h_t$ coupled with an observable $m(\underline
S(t))$, function of the system configuration $\underline S(t)$. The
Hamiltonian in presence of the field is
\begin{eqnarray}
H_{h}(\underline S(t))=H(\underline S(t))-h_t m(\underline S(t)). 
\end{eqnarray}
The response function is defined as usual 
\begin{eqnarray}
  \Delta (t,s)=\frac{\partial \langle m(\underline S(t))\rangle}{\partial
    h_s} 
\end{eqnarray} 
where the average is done over the multiple realizations of the
trajectories of the system. Because of the causal structure
of the Markov Chain (\ref{BPD}), the response function is non zero
only if $t>s$.
To analyse this quantity we start from
\begin{eqnarray}
\frac{\partial}{\partial h_s}P\left(\underline S(t),\underline S(t-1),...,\underline S(1)|\underline S(0)\right)=
\beta\left(m(\uS(s))-\E\big[m(\uS(s))|\underline S(s-1)\big]\right) P\left(\underline S(t),\underline S(t-1),...,\underline S(1)|\underline S(0)\right)
\end{eqnarray}
where 
\begin{eqnarray}
\E\left[m(\uS(s))|\underline S(s-1)\right]=\frac{1}{Z[\b;
  \underline S(s-1)]}\sum_{S'} e^{-\beta H(\underline S')} m(\underline
S') \delta\left(q(\underline S',\underline S(s-1))-\widetilde
  C(s,s-1)\right).  
\end{eqnarray}
This leads to 
\begin{eqnarray}
 \Delta (t,s)
=\beta[\left\langle m(\underline S(t))m(\underline S(s))\right\rangle-\langle
m(\uS(t))\E[m(\uS(s))|\underline S(s-1))\rangle] = \b[C(t,s)- D(t,s)] 
\end{eqnarray}
where 
\begin{eqnarray}
&&\left<A(\underline S(s)) \right>=\sum_{\underline
  S(t)}\ldots\sum_{\underline S(0)} A(\underline S(s))
P\left(\underline S(t),\underline S(t-1),...,\underline
  S(1)|\underline S(0)\right)\hat P(\underline S(0))\:,
\nonumber\\
&& C(t,s)=\left<m(\underline S(t))m(\underline S(s))\right>, \;\;\;\;
D(t,s)=\langle m(t)E(m(S'(s))|\underline S({s-1}))\rangle=
Q_{s-1}(t,s).
\label{clone}
\end{eqnarray}
Notice that, by simple properties of conditional probability, if $t<s$
one has $D(t,s)=C(t,s)$ and the response is
$\Delta(t,s)=0$  as it should be. Eq. (\ref{clone}) provides the announced relation
between response and clone correlation: a response at
large times is only possible if $Q_{s-1}(t,s)=D(t,s)$ differs from $C(t,s)$.

We would like now to make some remarks on the long chain (i.e.\ long times) limit.
By properly choosing the constraining value close to the typical overlap value, $\widetilde{C} \approx q_{EA}$, all the fast dynamics is coarse-grained in a single step of the pseudodynamics.
Consequently, we have $C(t,t) \approx q_{EA}$
and the equal time response $\Delta(t,t)=\beta[\langle m(t)^2 \rangle
-\langle m(t) \rangle^2]$, corresponding to the equilibrium-like response in the beta regime, is expected to be a quantity of order
one, which coincides with $\beta[C(t,t)-C(t+1,t)]$ since $D(t,t)$
should be close to $C(t+1,t)$. On the other hand if $t>s$ (to be intended as
$t\gg s$ in real times) one can expect $D(t,s)$ to be close to $C(t,s)$.  For a
chain of length $t$ the total susceptibility to a field that
acts from time $1$ to $t$, is $\chi(t)=\sum_{s=1}^t\Delta(t,s)$. This
quantity remains finite for $t\to\infty$ if
$\Delta(t,s)=R(t,s)\de s$ is infinitesimal in the continuous time limit. In this way the sum converges to 
\begin{eqnarray}
  \label{eq:risp}
  \chi(t)=\Delta(t,t)+\int_0^t \de s\; R(t,s)\;,
\end{eqnarray}
where the first and second terms are respectively the responses from the fast and the slow dynamics.
A non trivial response in the long time regime is thus associated with a non
zero response function $R(t,s)$ i.e. to decaying clone correlation
function that in the continuum limit is $Q_s(t,s+ds)=C(t,s)-\de s\, T\, R(t,s)$, since in the fast dynamics the fluctuation-dissipation relation $\partial_s C(t,s) = T\,R(t,s)$ holds.

\subsection{Equilibrium measure}
Despite we are interested to the applications of (\ref{BPD}) to glassy
dynamics and time scales where equilibration does not occur, in
general, for time independent 
correlations $\widetilde C(s+1,s)=\widetilde C$ and finite system's volumes, the
Markov chain (\ref{BPD}) is ergodic and it is interesting to study its
equilibrium measure. This is not the  ordinary Boltzmann
distribution. In fact we can observe that
the detailed balance is verified with respect to the modified
distribution
\begin{eqnarray}
\label{eq:mod}
\mu(S)=\frac{1}{Z_2}e^{-\beta H(\underline S)}Z(\beta, \underline S)
\end{eqnarray}
where 
\begin{eqnarray}
\label{eq:z2}
Z_2=\sum_{\underline S,\underline S'} e^{-\beta[H(\underline S)+H(\underline S')]} \delta\left(q(\underline S,\underline S')-\widetilde C\right)
\end{eqnarray}
This is therefore the equilibrium distribution of the
chain. Equivalently one can see that the measure for two configurations at consecutive times is
\begin{eqnarray}
  \label{eq:1}
  \mu_2(\underline S,\underline{S'})=\frac{1}{Z_2}e^{-\beta[H(\underline S)+H(\underline S')]} \delta\left(q(\underline S,\underline S')-\widetilde C\right).
\end{eqnarray}
\section{Mean-field glassy dynamics} \label{p-spin-without-replicas}
In this section we would like to analyse mean field spin glasses and
show how to obtain a full characterization of the dynamics in terms of
a single correlation function and its conjugated response
function. The analysis was previously performed in \cite{FP13} with
the aid of replica method that can be employed to treat analytically the denominators in
(\ref{BPD}). Here we propose an alternative derivation that avoids the
use of replicas similar to the derivation of mean-field dynamical
equations for Langevin dynamics presented in \cite{F07}.

We consider specifically the spherical $p$-spin model which provides the
canonical example of mean-field glassy dynamics.  The 
Hamiltonian of the model is 
\beq 
H_J[\underline S ;h]=-\sum_{i_1<\ldots <
  i_p}J_{i_1,\ldots,i_p}S_{i_1}\ldots S_{i_p}-\sum_{i=1}^N
h_i S_i\ \ \ \ \sum_{i=1}^NS_i^2=N\ \ \ \ P[J_{i_1\ldots
  i_p}]\propto \exp\left[-\frac{N^{p-1}}{p!}J_{i_1\ldots
    i_p}^2\right]\label{pspin} \eeq 
where we have introduced a site dependent  magnetic field
in the system that is needed in order to compute the
response function. In order to make the presentation as simple as possible we will restrict our
analysis to the case $p=3$ even if the general case is  a straightforward
generalization of this one. 

Due to the mean field nature of the model, we can get closed dynamical
equations in terms of two point correlation and response
functions. Let us consider an
arbitrary function or operator $\phi({\bf S})$ dependent on a
trajectory ${\bf S} =\{ {\underline{S}}(u)\}_{u=0}^{\tau}$ and write the
obvious identity
\begin{eqnarray}
  \label{eq:der}
\E_J  \int \de {\bf S}\frac{\partial}{\partial S_i(s)} \left[
\phi({\bf S})P({\bf S}|\underline S(0))\right]=0
\end{eqnarray}
where, $\mathbb E_J $ represents the average over the disordered couplings. In order to obtain
equations for correlation and response functions it is enough to
consider the insertion of $\phi(S)=S_i(\tau)$ and
$\phi(S)=\frac{\delta}{\delta h_i(\tau)}$, which do not depend on the
quenched variables $J$, we can therefore average direcly the measure 
$P({\bf S}|\underline S(0))$
\begin{eqnarray}
\label{mii1}
\mathbb E_J \frac{\partial}{\partial S_i(\sigma)} 
P({\bf S}|\underline S(0))=&
&\E_J\left\{ \left[\frac 1 2 \beta \sum_{j,k}J_{ijk}S_{j}(\sigma)S_{k}(\sigma)+\mu_\sigma S_i(\sigma)
\right.
\right.
\\
\nonumber
&&
\left.
\left.
+\nu_\sigma  S_i(\sigma-1)
+\nu_{\sigma+1} [S_i(\sigma+1)-
{\mathbb E}\left (S_i(\sigma+1)|{\bf S}(\sigma)\right)]\right]P({\bf S}|\underline S(0))\right\}.
\end{eqnarray}
In order to simplify our analysis we suppose that $\underline S(0)$ is chosen
randomly with uniform probability on the sphere and $P(\underline S(0))$ is
independent of $J$.  In the case where $P(\underline S(0))$ depends of $J$
additional terms would appear (see e.g. \cite{F07}). We can now
integrate by part the $J_{ijk}$ in the first term of (\ref{mii1})
which results in the substitution $J_{ijk}\to \frac{3}{N^2}
\sum_\epsilon \beta [
S_i(\epsilon)S_j(\epsilon)S_k(\epsilon)-\E(S_i(\epsilon)S_j(\epsilon)S_k(\epsilon)|{\bf
  S}(\epsilon-1))]$ to get
\begin{eqnarray}
\label{mii2}
\E_J \frac{\partial}{\partial S_i(\sigma)} 
P({\bf S}|\underline S(0))=&
&\mathbb E_J\left\{ \left[\frac {3} {2N^2} \beta \sum_\epsilon \beta \sum_{j,k}
[S_i(\epsilon)S_j(\epsilon)S_k(\epsilon)-\E(S_i(\epsilon)S_j(\epsilon)S_k(\epsilon)|{\bf S}(\epsilon-1))]
S_{j}(\sigma)S_{k}(\sigma)
\right.
\right.
\nonumber
\\
&&
\left.
\left.
+\mu_\sigma S_i(\sigma)
+\nu_\sigma  S_i(\sigma-1)
+\nu_{\sigma+1} [S_i(\sigma+1)-
\E\left (S_i(\sigma+1)|{\bf S}(\sigma)\right)]\right]P({\bf S}|\underline S(0))\right\}.
\end{eqnarray}
We will make at this point the crucial hypothesis that for typical
trajectories $\frac 1 N \sum_j S_i(\tau)S_i(\sigma)$ and $\frac 1 N
\sum_i S_i(\tau)\mathbb E(S_i(\sigma)|{\bf S}(\sigma-1))$ are self-averaging
quantities and coincide respectively with $C(\tau,\sigma)$ and
$D(\tau,\sigma)$. This hypothesis is equivalent to the factorization,
for $i\ne j$, $\langle
S_i(\tau)S_i(\sigma)S_j(\tau)S_j(\sigma)\rangle=\langle
S_i(\tau)S_i(\sigma)\rangle \langle S_j(\tau)S_j(\sigma)\rangle$ and
analogous formulas for more then two indices. This ``clustering
conditions'' of the correlation functions, which imply that the
averages are dominated by a single pure state, exclude a-priori
replica symmetry breaking (RSB) effects. It is of course possible to
include RSB effects even if in
the long chain limit, RSB effects are indistinguishable from violations
of FDT within the RS formalism.  Using the factorization hypothesis
 we finally get
\begin{eqnarray}
\label{mii2}
\E_J \frac{\partial}{\partial S_i(\sigma)} 
P({\bf S}|\underline S(0))=&
&\E_J\left\{ \left[\frac {3} {2} \beta \sum_\epsilon \beta
[S_i(\epsilon)C(\epsilon,\sigma)^2 -\E(S_i(\epsilon)|{\bf S}(\epsilon-1))D(\epsilon,\sigma)^2]
\right.
\right.
\nonumber
\\
&&
\left.
\left.
+\mu_\sigma S_i(\sigma)
+\nu_\sigma  S_i(\sigma-1)
+\nu_{\sigma+1} [S_i(\sigma+1)-
\E\left (S_i(\sigma+1)|{\bf S}(\sigma)\right)]\right]
P({\bf S}|\underline S(0))\right\}.
\end{eqnarray}
Inserting at this point $\phi(S)=S_i(\tau)$ and
$\phi(S)=\frac{\delta}{\delta h_i(\tau)}$ and performing the sum over the
trajectories we get respectively: 
\begin{eqnarray}
  \label{eq:eqC}
  -\delta(\tau-\sigma)=
  &&\sum_{\epsilon=0}^\tau \beta [ C(\tau,\epsilon)\; C(\sigma,\epsilon)^2-
D(\tau,\epsilon)\; D(\sigma,\epsilon)^2]
  +\mu_\sigma  C(\tau,\sigma)  +\nu_\sigma  C(\tau,\sigma-1)
  +\nu_{\sigma+1} \Delta(\tau,\sigma+1)\;,
\\
  \label{eq:eqR}
0=
  &&\sum_{\epsilon=\tau}^\sigma \beta \Delta(\epsilon,\tau)\; 
[C(\sigma,\epsilon)^2-D(\sigma,\epsilon)^2]
  +\mu_\sigma  \Delta(\tau,\sigma) 
  +\nu_\sigma  \Delta(\tau,\sigma-1)
  +\delta_{\tau,\sigma+1}\nu_{\sigma+1} \Delta(\sigma+1,\sigma+1)\;. 
\end{eqnarray}
These equations have a causal structure that is inherited from the chain
construction and can be integrated iteratively step by step. In the next section
we show that assuming the existence of a long chain limit, these
equations reduce to the long time equations for the slow part of the
Langevin dynamics of the same model. The consistency of this limit can be
checked through explicit integration of the equations. At low
temperatures, one finds an aging regime where 
selfconsistently  $\widetilde{C}(\tau,\tau+1)\to q_{EA}$ while $\nu_\tau
\to 0$ at large $\tau$. 

\subsection{The long chain limit}

The equations (\ref{eq:eqC},\ref{eq:eqR}) can be easily generalized to arbitrary
spherical long range spin glass models with Hamiltonian of $p$-spin
mixture type $H[\underline S]=\sum_{p}a_p H_p[\underline S]$ and 
correlation function $E(H(\underline S)H(\underline S'))=Nf(\underline S\cdot \underline S'/N)$, with $f(q)=\frac
1 2 \sum_p a_p^2 q^p$.  It is interesting to write them in the general case: 
\begin{eqnarray}
  \label{eq:eqC1}
  -\delta(\tau-\sigma)=
  &&\sum_{\epsilon=0}^\tau \beta[ C(\tau,\epsilon)\; f'(C(\sigma,\epsilon)-
D(\tau,\epsilon)\; f'(D(\sigma,\epsilon))]
  +\mu_\sigma  C(\tau,\sigma)  +\nu_\sigma  C(\tau,\sigma-1)
  +\nu_{\sigma+1} \Delta(\tau,\sigma+1)
\\
0=
  &&\sum_{\epsilon=\tau}^\sigma \beta \Delta(\epsilon,\tau)\; 
[f'(C(\sigma,\epsilon))-f'(D(\sigma,\epsilon))]
  +\mu_\sigma  \Delta(\tau,\sigma) 
  +\nu_\sigma  \Delta(\tau,\sigma-1)
  +\delta_{\tau,\sigma+1}\nu_{\sigma+1} \Delta(\sigma+1,\sigma+1). 
\end{eqnarray}

In the continuum time limit, using $D(t,s)=C(t,s)-T\, R(t,s)ds$ for
$t>s$, and fixing $\nu_t=0$, that corresponds to using the saddle-point value $q_{EA}$ for $\widetilde C(t,t-1)$, we get 
\begin{eqnarray}
\label{cont}
\mu(t) C(t,u)
= &&\beta \int_0^u \de s\; 
f'(C(t,s)) R(u,s)+ \beta \int_0^t \de s\; f''(C(t,s)) R(t,s)C(u,s)
\nonumber\\ 
&& +\beta^2 (f'(1)-f'(q_{EA}))C(t,u)+\beta^2
f'(C(t,u))(1-q_{EA}) \\\nonumber
% && +\beta \beta f'(C(t,0))C(u,0),\\\nonumber
\mu(t) R(t,u)
= & &
\beta\int_u^t \de s f''(C(t,s))R(t,s) R(s,u)\\ \nonumber
  &&+\beta f''(C(t,u))R(t,u)(1-q_{EA})
%\\\nonumber
% &&
+\beta (f'(1)-f'(q_{EA}))R(t,u),\\ \nonumber
  \mu(t)=&&
 T +\beta^2(f'(1)-f'(q_{EA}))\\ 
 &&+\beta \int_0^t \de s\; 
\left( f'(C(t,s)) R(t,s)+ f''(C(t,s)) R(t,s)C(t,s)\right)
%+\beta^2
%'(C(t,0))C(t,0),
 \nonumber\\
\nonumber
\end{eqnarray}
where we used the condition ${\widetilde C}(t,t)=1$ and
$C(t,t)=q_{EA}$. These are well know equations that in the dynamic
theory of  mean field spin glasses, depending on the models, they have
a dynamical phase transition below which  there are aging solutions
where fluctuation dissipation theorem and time translation
invariance do not hold. 
 
\begin{figure}[!t]
  \includegraphics[width=.7 \textwidth] {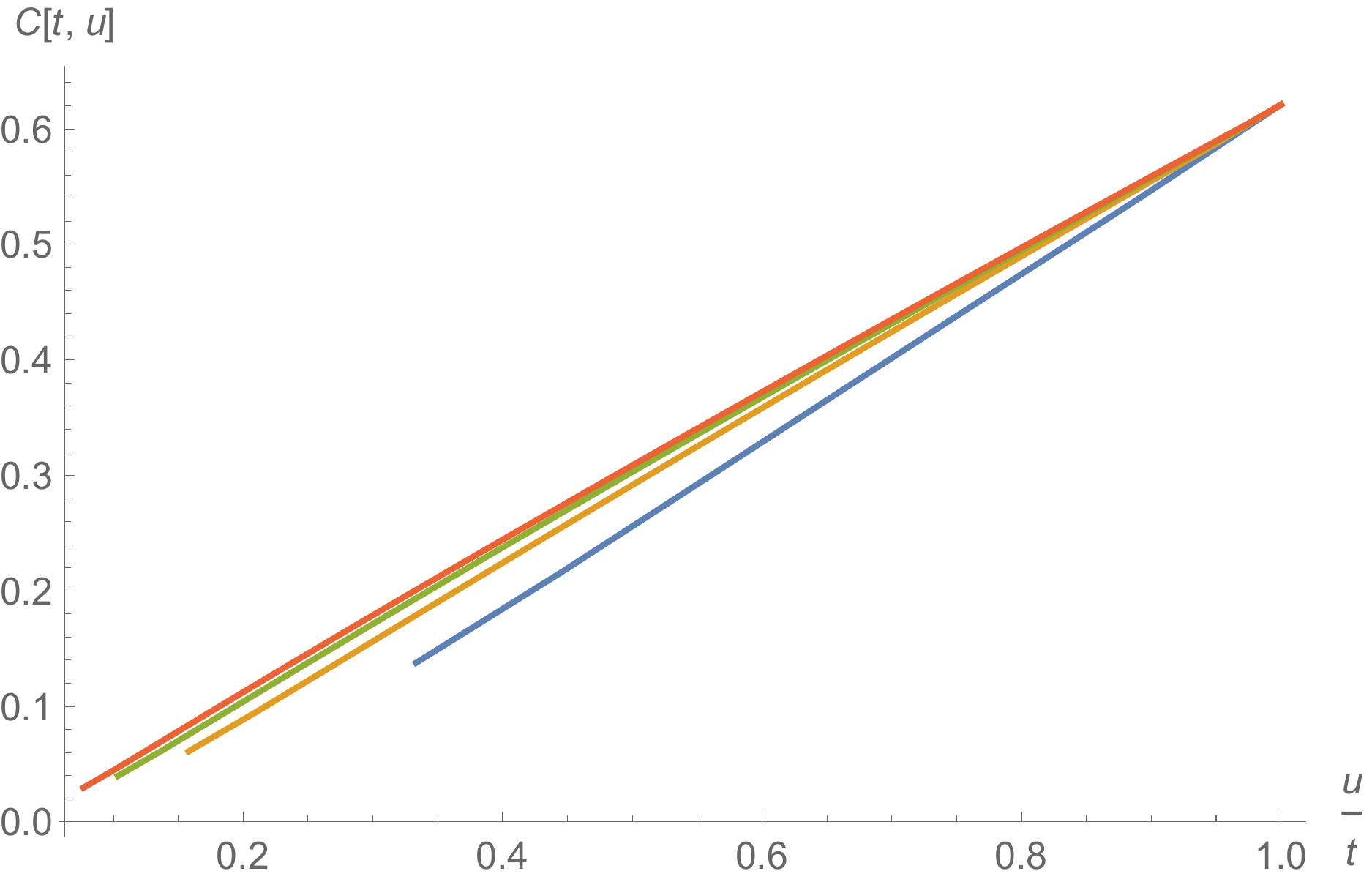}
\caption{The correlation function $C(t,u)$ in the spherical $p$-spin
  model for $p=3$ from the explicit numerical
  integration of the equations as a function of $u/t<1$ for
  $t=9 (blue), 19 (orange), 29 (green), 39 (red)$. The integration is performed at $T=0.517< T_d=\sqrt{3/8}=0.612$. $C(t+1,t)$ is kept fix at the value
  $q_{EA}=0.621$. The data seem to indicate that $C(t,u)$ tends to an
  exclusive function of the ratio $u/t$ for large $u$ and $t$.}
  \label{fig1}
\end{figure}

In order to check the consistency of the aging solutions and the long
time limit of the original Markov Chain, we have integrated explicitly
the discrete BPD equations for the pure $p$-spin model for $p=3$ with
$f(q)=\frac 1 2 q^3$ and for a mixture of $p=2$ and $p=4$ with
$f(q)=\frac 1 2 q^2+\frac{1}{20} q^4$.  It is interesting to consider
both models since, as well known, the former has a one-step
replica symmetry breaking (1RSB) phase and displays aging with a single
(inverse) effective temperature \cite{CK93} $\beta_{eff}=\beta X$ while the
latter has a full replica symmetry breaking (fRSB) phase and has a
continuum set of effective temperatures $\beta_{eff}(q)=\beta
X(q)=f'''(q)/f''(q)^{3/2}$ that during aging depend continuously on
the value of the correlation function $C(t,u)=q$. We integrated the
equations in the low temperature regime with $\widetilde{C}(t,t-1)$ fixed
to the theoretical value of $q_{EA}$ found from self-consistency of
the long time equations with $\nu_t=0$. An example of the results obtained from the numerical
integration of the equations is given in Fig. \ref{fig1} where we show the
correlation function of the $p$-spin model for $p=3$ at a low
temperature.  A popular way of analyzing aging systems and detecting
effective temperatures consists in looking at the parametric plot of
the integrated response function
$T\chi(t,s)=\sum_{\epsilon=\sigma}^\tau \Delta(\tau,\epsilon)\to
T\int_s^t du \; R(t,u)$,  as a function of the correlation function
$C(t,s)$. According to mean-field theory, for large times these curves
tend to a master curve whose slope should be precisely $-X(C)$.

\begin{figure}[!t]
\includegraphics[width=.7 \textwidth]{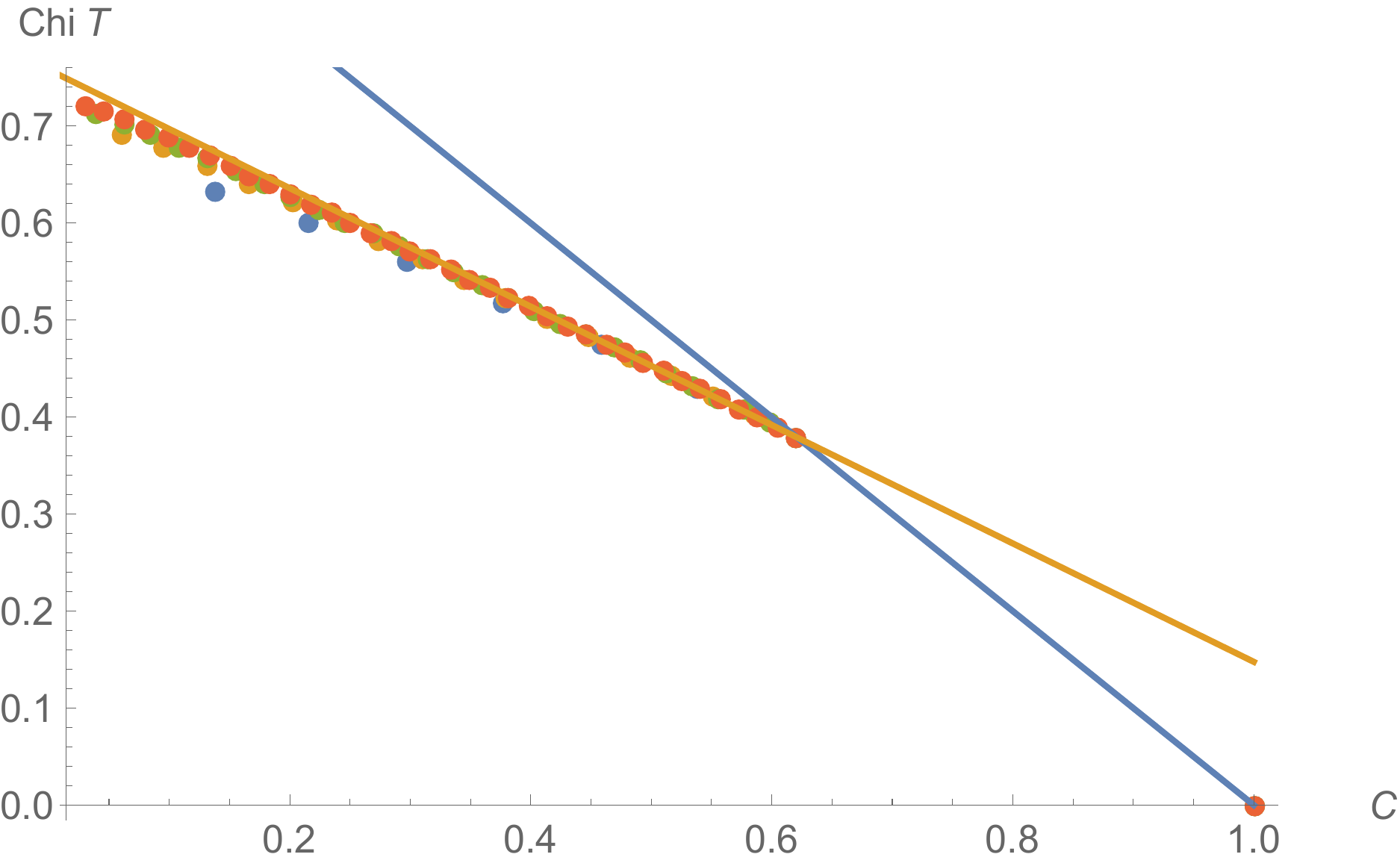}
\caption{The plot of rescaled susceptibility $T\chi(t,u)$ as a
  function of $C(t,u)$ in the $p$-spin model for $p=3$. The lines
  are computed at times $t=9 (blue), 19 (orange), 29 (green), 39 (r ed)$
  and the temperature is $T=0.517$ ($T_d=\sqrt{3/8}=0.612$). We also
  plot the FDT line $1-C$ and the modified FD prediction
  $1-q_{EA}+X(q_{EA}-C)$. The value of the fluctuation-dissipation
  ratio predicted by the long time dynamics of Langevin equation is
  $X=0.610$ and coincides well with the data. }
\label{fig2}
\end{figure}

\begin{figure}[!t]
\includegraphics[width=.7 \textwidth]{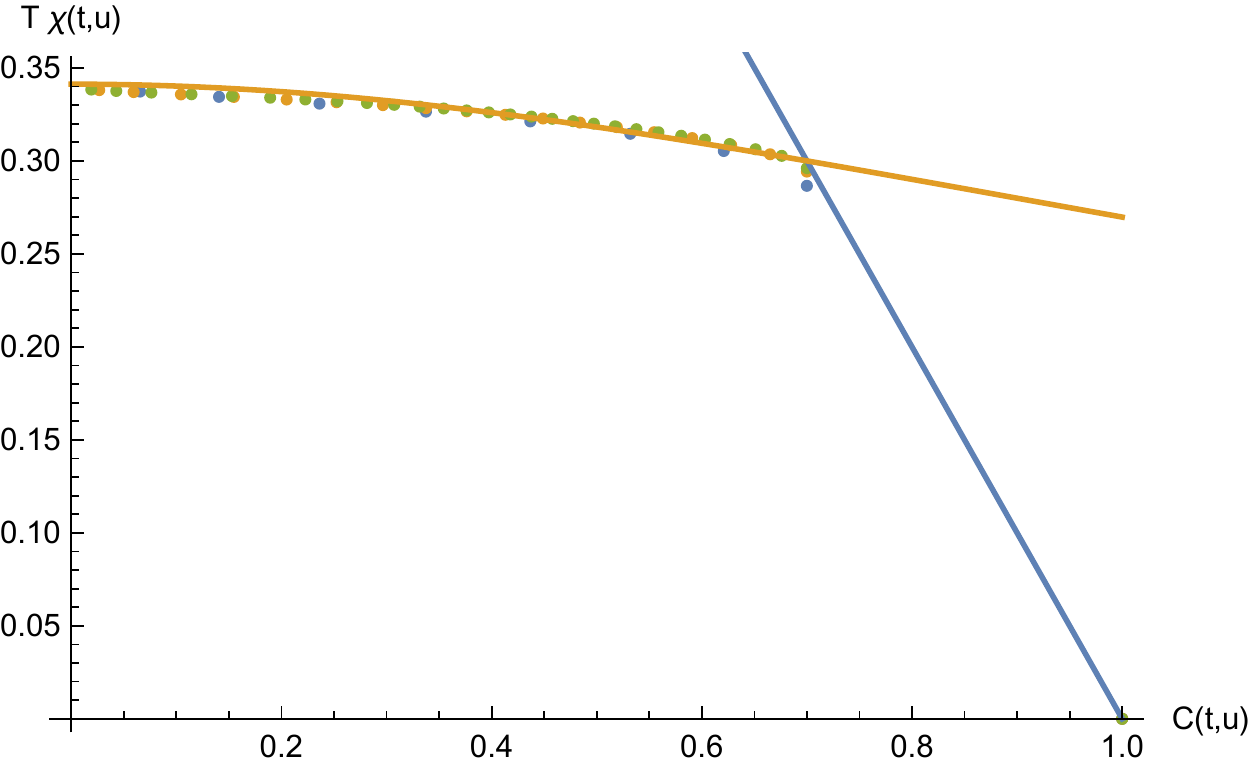}
\caption{The plot of the rescaled susceptibility $T\chi(t,u)$ as a
  function of  $C(t,u)$ for $t=10 (blue), 20 (orange), 30 (green)$ in the 2+4 model with
  $f(q)=\frac 1 2 (q^2 +a q^4)$ and $a=0.1$ at temperature $T=0.34$.  The critical temperature
 of the model is $T_c=1$ . We also plot the FDT
  line $1-C$ and the modified FD prediction from RSB, which show the
  good agreement of the BPD with the expected results for infinite
  time prediction.  
}
\label{fig3}
\end{figure}

In Fig.s \ref{fig2} and \ref{fig3} we present such plot
respectively for the $p=3$ model and for the $2+4$ model, and compare
it with the asymptotic result predicted by the aging mean field
theory. We see that in both the 1RSB and fRSB cases, the asymptotic
aging limit is approached after a very limited number of steps in the
chain. We confirm in this way that BPD gives a faithful representation
of aging dynamics.

\section{Replicas}

In the previous section we used a path-integral representation of the
probability of a trajectory to get a set of closed equations for the correlation
and response functions. In the case of systems with long range
interactions an alternative way to produce the dynamical equations is by treating the denominators that
appear in the definition (\ref{BPD}) of the Boltzmann Markov Chain with the replica method. This consists
in substituting in all the theoretical computations the denominator
appearing in (\ref{BPD}) by a positive power of the constrained
partition function to get 
\begin{eqnarray}
\label{Mrep}
M_{n_{t+1}}
(\underline S (t+1)|\underline S(t))={Z[\beta; \underline
  S(t)]^{n_{t+1}-1}}\mathrm{e}^{-\beta H[\underline
  S(t+1)]}\delta\left(\widetilde C(t,t+1)-q\left(\underline
    S(t),\underline S({t+1})\right)\right) 
\end{eqnarray}
The new chain depends on the parameters $n_{t+1}$ that for each time
$t$ counts the ``number of replicas''. As usual in the replica method,
these number are
considered integers in intermediate computations, but sent to zero 
in the replica limit where (\ref{Mrep}) coincides with
(\ref{BPD}). Renaming the 
``master replica'' $S(t)=S^0(t)$ and introducing $n_t-1$ ``slave replicas''
$S^a(t)$, $a=1,...,n_t-1$, we get:
\begin{eqnarray}
  M_{n_{t+1}} (\underline S^0 (t+1)|\underline S^0(t))=\sum_{\{S^a(t+1)\}_{a=1}^{n_{t+1}-1}}
  \mathrm{e}^{-\beta\sum_{a=0}^{n_{t+1}-1} H[\underline
    S^a(t+1)]} \prod_{a=0}^{n_{t+1}-1} \delta\left(\widetilde
    C(t,t+1)-q\left(\underline S^0(t),\underline
      S^a({t+1})\right)\right)
\end{eqnarray}
The dynamical equations for the correlations and response functions in the case of the $p$-spin spherical model
(\ref{cont}) were first derived with this formalism in \cite{FP13}. 
Within the replica method it is natural to introduce the total partition
function up to time $t$
\begin{eqnarray}
  \label{eq:3}
  Z_{tot}(t)=\sum_{S^a(u)} e^{-\beta \sum_{u,a} H[S^a_t]}\prod_{u,a}\delta(q(S^a(u),S^0(u-1))-C(u,u-1))
\end{eqnarray}
where in the sum and the product $u$ runs up to $t$ and, for each $u$, 
$a$ runs from $0$ to $n_u-1$. This expression, complicated as it may
be, has the same formal structure of a partition function of a
replicated equilibrium system. The difference with the usual case
comes in the fact that here we have an explicit interaction between
replicas at subsequent times due to the chain constraint.  As in the
usual case however, the finite $n_t$ system can be interpreted as a
mixture of interacting particles of different kind.  In this way this is a
good starting point for approximations. In fact one can apply quite
straightforwardly all existing
approximations for equilibrium mixtures modulo parametrizations of the
quantities of interest that allow for the analytic continuation needed to take the
replica limit. As usual, considerations of symmetry under permutations
of different replicas play an important role.  In the present case,
differently from the equilibrium case where all replicas are equivalent,
the partition function (\ref{eq:3}) is only invariant under 
independent permutations of groups of slave replicas with the same time
index.

In replica calculations it is quite simple to see that a prominent role in manipulating (\ref{eq:3}) is played by the replica correlation function
$Q_{ab}(s,u)=\langle S^a(s)S^b(u)\rangle$. This codes for both the
correlation and the response functions of the previous
section. Assuming replica symmetry (i.e. invariance under the
previously mentioned group of permutations), we can write the most generic replica symmetric parametrization for the overlap matrix
\begin{eqnarray}
  \label{eq:11}
   Q_{ab}(s, u) = C(s, u) +  \delta_{b,1} \Delta(s,u)
   +\delta_{a,1}\Delta(u,s) +\delta_{s,u} \delta_{a,b}\Delta(u,u). 
\end{eqnarray}
In the
long chain continuous time limit this becomes:
\begin{eqnarray}
  \label{eq:4}
  Q_{ab}(s, u) = C(s, u) + R(s,u)\de u\,\theta(s-u) \delta_{b,1}
+ R(u,s)\de s\,\theta(s-u) \delta_{a,1}  + [\widetilde{C}(u,u) - C(u, u)]\delta_{s,u}\delta_{a,b}.
\end{eqnarray}
In \cite{FP13} it has been shown that this parametrization reproduces the correct dynamical equations for this model in the slow dynamical regime. 
In the next section we will deeply use this form in the context of the replicated liquid theory in order to obtain a set of new dynamical equations for structural glasses \cite{FPU15}.

\section{Boltzmann Pseudodynamics for supercooled liquids}
In this section we want to develop the BPD formalism to describe the slow regime of the dynamics of supercooled
liquids undergoing to a glass transition.
As we emphasized in the previous section this can be done quite easily.
Indeed the BPD construction can be directly applied to the replicated liquid theory that has been successfully employed to describe the glass transition \cite{PZ10, MP96, CFP98, CFP98b, FJPUZ12PNAS, FJPUZ13JChemPhys}.
In particular we can take the equations of the replicated liquid theory used to obtain the statics of structural glasses
and plug the pseudo dynamics ansatz inside them.
Practically, in doing this we have to promote simple replica indices to BPD replica indices $\a\to(t,a)$.

%
%In this section we use the replicated BPD in order to get an approximate
%descriptions of slow glassy dynamics of supercooled liquids.  As
%stessed in the previous section, for many respects, the formal
%structure of BPD of liquid systems with replicated denominators is
%similar to the one of replicated atomic systems used to study
%equilibrium and modified equilibrium of glasses.  This becomes
%particularly clear if the time and replica indexes $(t,a)$ are
%composed into a unique index $\alpha$. The difference with the
%equilibrium approach of glasses comes in the fact that while in this
%last case one has a number of replicas $n$ which are treated
%symmetrically with $n\to 1$ at the end, in the BPD one has that the
%replicas are constrained to be in a chain. While this is of
%course important at the end of the computations when replica indexes
%are disentagled in order to get the wanted
%dynamical description, it is irrelevant  at the initial stages of the
%computations. 
 Let us see how all this procedure works.
We start from the definition of the basic quantities that can be treated in the theory of the replicated liquid \cite{MP96}. The simplest objects we need are the density field and its two point correlation function that are defined by
\beq
\r_\alpha(x)=\langle \sum_{i=1}^N\delta(x-x_i^{(\alpha)})\rangle\ \ \ \ \ \ \ \ \ \r_{\alpha\beta}(x;y)=\langle\sum_{[ij]}\delta(x-x_i^{(\alpha)})\delta(y-x_j^{(\beta)})\rangle
\eeq
where the sum over $[ij]$ runs on all $i,j$ if $\alpha\neq \beta$ and over $i\neq j$ if $\alpha=\beta$. Moreover we define
\beq
h_{\alpha\beta}(x,y)=\frac{\r_{\alpha\beta}(x,y)}{\r_\alpha(x)\r_\beta(y)}-1\:;
\eeq
in what follows we will always look for a uniform solution for the density field such that $\r_\alpha(x)=\r$. 

We introduce also the direct correlation function $c_{\alpha\beta}(x)$ that is defined by the 
replicated Orstein-Zernike (OZ) equation: 
\begin{gather} 
c_{\alpha\beta}(x)
=h_{\alpha\beta}(x)-\sum_{\gamma=1}^{n_{tot}} \int \de y\; h_{\alpha\gamma}(x-y)\r\,
  c_{\beta\gamma}(y)\label{OZ}. 
\end{gather} 
It is convenient also to rewrite the same equations in Fourier space
\begin{gather} 
c_{\alpha\beta}(q)=h_{\alpha\beta}(q)-\sum_{\gamma=1}^{n_{tot}} h_{\alpha\gamma}(q)\r\, 
  c_{\beta\gamma}(q)\label{OZ}. 
\end{gather} 
In the glass phase and in within a 1RSB picture, the static version of the OZ equation can be written as
\begin{gather}
\tilde h(q)=\tilde c(q)+\r\left[\tilde h(q)\tilde c(q)+(m-1)h(q)c(q)\right]\label{OZ1am}\\
h(q)=c(q)+\r\left[\tilde h(q)c(q)+\tilde c(q)h(q)+(m-2)h(q)c(q)\right]\label{OZ2am}
\end{gather}
where $m$ is the number of replicas. In the limit $m\to 1$ we get the OZ equations that are needed to compute the dynamical MCT transition point
\begin{gather}
\tilde h(q)=\tilde c(q)+\r\,\tilde h(q)\tilde c(q)\label{OZ1}\\
h(q)=c(q)+\r\left[\tilde h(q)c(q)+\tilde c(q)h(q)-h(q)c(q)\right]\label{OZ2}
\end{gather}
where $\tilde c$ and $c$ are respectively the diagonal and off-diagonal parts of the matrix $c_{\a\b}$ and the same is true for $h_{\a\b}$.
These quantities can be thought as the corresponding correlations functions for the supercooled liquid for times that are such that any dynamical correlation function is close to its plateau value. In particular they can be used to compute the MCT non ergodicity parameter.

We want to show that once the relevant generalization of
(\ref{eq:4}) is considered for $c_{\alpha\beta}$ and $h_{\alpha\beta}$, the equation
(\ref{OZ}) has formally the structure of a long time 
Mode Coupling equation which can be used both to describe equilibrium
slowing down as the glass transition is approached and the aging dynamics
below the glass transition point. 

As in the simple case of (\ref{eq:4}) where there is no
space dependence,  the replica
dependence of $h$ and $c$ encode for correlation and response functions and
\begin{gather} 
h_{\alpha\beta}(x)=h_{ab}(s,u,x)=
h(s,u;x)+\delta_{ab}\delta_{su}\Delta
h(s,s;x)+
\delta_{a1}
R_h(u,s;x)\de s
+\delta_{b1}
R_h(s,u;x)\de u
\label{FP1}\\
c_{\alpha\beta}(x)=c_{ab}(s,u,x)=c(s,u;x)+\delta_{ab}\delta_{su}\Delta
c(s,s;x)+
\delta_{a1}
R_c(u,s;x)\de s
+\delta_{b1}R_c(s,u;x)\de u.
\label{FP2}
\end{gather}
Plugging these forms inside Eq. (\ref{OZ}), we get
a dynamical version of the OZ equations
\begin{eqnarray}
\label{OZvetri}
&&h(q;s,u)=c(q;s,u)+\r\left[h(q;s,0)c(q;0,u) +h(q;s,u)\Delta c(q;u,u)+\Delta h(q;s,s) c(q;s,u)+\right.\\
&&\left.+\frac 1\beta \int_0^u \de z\, h(q;s,z)R_c(q;u,z)+
\frac 1\beta \int_0^s \de z\, R_h(q;s,z)c(q;z,u)\right]\\
&&\Delta h(q;s,s)=\Delta c(q;s,s)+\r\Delta h(q;s,s)\Delta c(q;s,s)\\
&&R_h(q,s,u)=R_c(q;s,u)+\r\left[R_h(q;s,u)\D c(q;s,s)+\D h(q;u,u)R_c(q;s,u)+\frac 1\b \int_u^s\de z\, R_h(q;z,u)R_c(q;s,z)\right]
\label{OZvetri_fin}
\end{eqnarray}
These equations are not closed: we need to provide some kind of closure scheme in order to have a complete set of dynamical equations.
This will be done in the following. For the moment, let us investigate the properties of these relations.
It is quite simple to see that these equations are
compatible with time translation invariance (TTI) and fluctuation
dissipation theorem (FDT):
\begin{eqnarray}
  \label{eq:5}
&&  h(q;s,u)=h(q;s-u)\nonumber\\
&& R_h(q;s,u)=\theta(s-u)\beta
  \frac{\partial}{\partial u}h(q;s-u), 
\end{eqnarray}
and analogous relations for  $c$ and $R_c$. If TTI and FDT are
inserted in the OZ equations, they reduce to the following equation (we can set $u=0$ due to TTI):
\begin{equation}
h(q;s)=c(q;s)+\r \left[h(q;s)\Delta c_0(q)+\Delta h_0(q)c(q;s)+\frac 1\beta \int_0^s \de z R_h(q;s-z)c(q;z)+h(q;s)c(q;0)\right]\label{hTTI}
\end{equation}
where we have introduced the following notation
\begin{gather}
\Delta h(q;s,s)=\Delta h_0(q)=\tilde{h}(q;0)-h(q;0)\ \ \ \ \Delta c(q;s,s)=\Delta c_0(q)=\tilde c(q;0)-c(q;0)
\end{gather} 
where $\tilde h(q;0)$ and $h(q;0)$ are nothing but a solution of the static OZ equations.

\subsection{Closure schemes}
The OZ relations alone are not enough to write a self-consistent system of equations and we need to provide a closure scheme for them.
Here we discuss two closure schemes: the first one is the standard Hypernetted Chain Approximation (HNC) \cite{Hansen} that has been developed extensively to study glasses \cite{PZ10, CFP98}. The second one is a closure scheme that has been introduced by G. Szamel \cite{Sz10} in order to derive the standard MCT equations for the non ergodicity parameter from the replica approach. 
Both the two approaches have advantages and disadvantages. On the one hand the HNC approximation is known to not provide a quantitative sensitive description of the glass transition \cite{FJPUZ13JChemPhys}. On the other it is variational since it can be derived from a partial resummation of the diagrams that give the free energy. This makes it quite suitable for a systematic improvements on top of it.
The Szamel's closure instead is an \emph{ad hoc} scheme to obtain quantitatively the same non-ergodicity factor of standard MCT from the replica approach. The disadvantage is that this procedure is not variational and it needs the external input of the static structure factor as it is usual in MCT.
However quite remarkably, using the BPD construction we are able to derive from these purely static approximation/closure schemes a set of dynamical equations that are nothing but MCT equations in the long time regime.

\subsubsection{Dynamical HNC equation}
The HNC closure equation for a replicated system is given by
\beq
\label{HNCclosure}
\begin{split} \ln[h_{\alpha\beta}(x,y) &+
1]+\beta\phi_{\alpha\beta}(x,y)=h_{\alpha\beta}(x,y)-c_{\alpha\beta}(x,y)
\end{split}
\eeq
where for us $\phi_{\alpha\beta}(x,y)\equiv
\phi_{a\tau;b\sigma}(x,y)
=\delta_{\tau,\sigma}\delta_{a,b}\phi(x-y)+\delta_{b,1}\delta_{\tau,\sigma+1}w(x-y)
\nu(\sigma)$ contains the inter particle potential at equal time and
replica indexes %%as well as %%Rimosso and %%Aggiunto the Lagrange
multiplier constraining the value of the overlap at consecutive times.
Plugging the parametrization (\ref{FP1}-\ref{FP2}) into
(\ref{HNCclosure}) we obtain for $s\ne u$
\begin{gather}
\label{hR} \ln[h(x;s,u)+1]=h(x;s,u)-c(x;s,u)\\
R_c(x;s,u)=R_h(x;s,u)\frac{h(x;s,u)}{h(x;s,u)+1}\:.\label{hR2}
\end{gather} 
The dynamical equations (\ref{OZvetri}-\ref{OZvetri_fin},\ref{hR},\ref{hR2}) provide a complete set of equations that
can be solved in time \cite{FPU15}. 

It is quite evident from the BPD construction that the equations that we can get from it must be covariant under time reparametrization.
Technically this means that if we have a solution $h(q;s,u)$ and $c(q;s,u)$ for these equations we can obtain another solution from this one in the following way:
we consider a monotonically increasing function $f(t)$; we can write a new solution as
\begin{gather}
h'(q;s,u)=h(q;f(s),f(u))\ \ \ c'(q;s,u)=c(q;f(s),f(u))\\
R_h'(q;s,u)=\frac{\de f(u)}{\de u}R_h(q;f(s),f(u))\ \ \ R_c'(q;s,u)=\frac{\de f(u)}{\de u}R_c(q;f(s),f(u))
\end{gather}
The striking consequence of this fact is that time here is just an arbitrary parameter.
Reparametrization invariance plays the role of a gauge symmetry: if we want to obtain physical observables we need to fix the gauge.
In this way we can reduce the degrees of freedom contained in the equations. 
A way to do it is to consider the equations in the regime where TTI and FDT hold.
If we impose both of these conditions we get a unique relation that is given by
\beq
\begin{split}
0&=c(q,s)-h(q,s)+\r \left[h(q;s)\Delta c_0(q)+\Delta h_0(q)c(q;s)-\int_0^s \de z\, \dot h(q;s-z)c(q,z)+ h(q;s)c(q;0)\right]\\
&=W_q[h]-\r\int_0^s \de z\, \dot h(q,z)[c(q,s-z)-c(q;s)]\label{HNC-MCT}
\end{split}
\eeq
where
\beq
W_q[h]=c(q;s)-h(q;s)+\r \left[h(q;s)\Delta c_0(q)+c(q;s)\Delta h_0(q)+c(q;0)h(q;s)-(h(q,s)-h(q,0))c(q,s)\right]\:.
\eeq
We immediately note that this equation is nothing but a MCT equation where the MCT kernel is replaced by the direct correlation function.
This has the consequence that different modes $q$ in the system are coupled as it should be since the direct correlation function can be expressed in terms of a series expansion in $h(q,t)$
\begin{gather}
c(q;t)=\sum_{n=2}^\infty \frac{(-1)^n}{n}\int \frac{\de^D k_1}{(2\pi)^D}\ldots \int \frac{\de^D k_{n-1}}{(2\pi)^D}h(k_1;t)\ldots h(k_{n-1},t) h(q-k_1-\ldots -k_{n-1};t)\:.
\end{gather}
From this equation we can obtain the mode-coupling exponent parameter $\l_{\textrm{MCT}}$. This quantity encodes for the dynamical exponents that characterize the approach and the departure of the dynamical correlation functions from their plateau value. The schematic way to obtain this quantity is to expand the dynamical equations using
\beq
h(q;t)=h(q;0)+G_q(t)\ \ \ \ \ \ \ \ \ \ \ G_q(t)=Ak_0(q)t^{b}+\delta G_q(t)
\eeq 
where $k_0(q)$ is the zero mode eigenvector \cite{FJPUZ12PNAS}.
In this way we get
\beq
\l_{\textrm{MCT}}\equiv \frac{\Gamma^2(1+b)}{\Gamma(1+2b)}=\frac{\int\de^D x\frac{k_0^3(x)}{(1+\tilde h(x))^2}}{2\r\int_q\,k_0^3(q)(1-\r\,\Delta c(q))^3}\:.
\eeq
This result has been derived also from a different perspective in \cite{FJPUZ12PNAS,FJPUZ13JChemPhys}.
Moreover we can also use the dynamical equations in the aging regime where we have done a quench of the system from a high temperature configuration down to a temperature lower than the MCT one. 
Because we are in the aging time window, we can set to zero the term $h(q;s,0)c(q;u,0)$ and the dynamical equations become
\beq
\begin{split}
h(q;s,u)&=c(q;s,u)+\r\left[ h(q;s,u)\D c(q)+\D h(q) c(q;s,u) +\frac{1}{\beta}\int_0^u \de z R_c(q;u,z)h(q;s,z)+\right.\\
&\left.+\frac 1\b \int_0^s \de z R_h(q;s,z)c(q;z,u) \right]
\end{split}
\eeq
\beq
R_h(q,s,u)=R_c(q;s,u)+\r\left[R_h(q;s,u)\D c(q)+\D h(q)R_c(q;s,u)+\frac 1\b \int_u^s\de z R_h(q;z,u)R_c(q;s,z)\right]\:.
\eeq
We can now consider the aging parametrization for the correlation functions
\beq
\begin{split}
h(q;s,u)&=\underline h\left(q;\frac us\right)\\
R_h(q;s,u)&=\frac 1s \mathcal R_h\left(q; \frac us\right)\\
c(q;s,u)&=\underline c\left(q;\frac us\right)\\
R_c(q;s,u)&=\frac 1s \mathcal R_c\left(q; \frac us\right)
\end{split}
\eeq
and setting $\l=u/s$, the equations become
\beq
\begin{split}
\underline h(q;\l)&=\underline c(q;\l)+\r\left[ \underline h(q;\l)\D c(q)+\D h(q) \underline c(q;\l) +\frac{1}{\beta}\int_0^\l \frac{\de \l'}{\l} \mathcal R_c\left(q;\frac{\l'}{\l}\right)\underline h(q;\l')+\right.\\
&\left.+\frac 1\b \int_0^1 \de \l' \mathcal R_h(q;\l')\underline c\left[q;\left(\frac{\l'}{\l}\right)^{\textrm{sgn}(\l-\l')}\right] \right]\label{Cageing}
\end{split}
\eeq
\beq
\mathcal R_h(q,\l)=\mathcal R_c(q;\l)+\r\left[\mathcal R_h(q;\l)\D c(q)+\D h(q)\mathcal R_c(q;\l)+\frac 1\b \int_\l^1\frac{\de \l'}{\l'} \mathcal R_h\left(q;\frac{\l}{\l'}\right)\mathcal R_c(q;\l')\right]\:.\label{Rageing}
\eeq
By using the quasi-fluctuation dissipation ansatz
\beq
\begin{split}\label{quasiFDT}
\mathcal R_h(q;\l)&= \b x \frac{\de}{\de \l}\underline h(q;\l)\\
\mathcal R_c(q;\l)&= \b x \frac{\de}{\de \l}\underline c(q;\l)\:.
\end{split}
\eeq
we can obtain that the value of $x$ is fixed by the marginal stability condition according to which the dynamical equations must have a zero mode that in replica theory is called the replicon \cite{CK93, FJPUZ12PNAS, FJPUZ13JChemPhys}.
In this way all the off-equilibrium dynamics follow closely the one of the $p$-spin glass model.

\subsubsection{MCT from BPD}
It has been shown in \cite{Sz10} how to construct a consistent truncation scheme of the replicated BBGKY hierarchy in order to obtain from replicas the equation of the non ergodicity parameter that has been derived within MCT. 
In what follows we want to go beyond the non-ergodicity factor to obtain the whole MCT dynamical equation in the long time limit. 
We can do this exactly on the same lines as we did in the HNC approximation scheme.
In this case the closure scheme is provided giving the non-diagonal elements of the replicated direct correlation function
 $c_{\alpha\neq \beta}(q)$
in terms of the static direct correlation function
$c_0(q)$: 
\begin{eqnarray}
  \label{eq:7}
c_{\alpha\beta}(k)=\int dq \; V(k,q) h_{\alpha\beta}(q)
h_{\alpha\beta}(k-q)
\label{sz}
\end{eqnarray}
where the  $V(k,q)$ is the Mode Coupling vertex function
\begin{eqnarray}
V(k,q)=\frac{1}{16\pi^3 k^2} [{\hat {\bf k}}\cdot ({\bf q}
c_0(q)+({\bf k-q})c_0(k-q))]^2,  
\end{eqnarray}
which is independent of the replica indexes.  
%Consistently with our previous comment, 		%%Rimosso
At this point we use again the mapping of replica indices on pseudo time indices 
$\alpha\to (a=1,t)$, $\beta\to (a=1,s=0)$ with $t>0$.
Within this scheme the direct static correlation function is supposed to come from equilibrium
and this is the only regime we can have access to.
By using TTI and FDT we get 
\begin{eqnarray}
  \label{eq:direct}
c(k,t)=\int dq \; V(k,q) h(q,t)h(k-q,t).   
\end{eqnarray}
Plugging this equation inside (\ref{hTTI}),
after some simple algebra, we get the MCT equations derived by G\"otze \cite{GoBook}.

\section{Conclusions}

In this paper we have reviewed the construction of the Boltzmann
pseudodynamics and presented some new results for spin glasses and
liquid theory. Among them we have: 
\begin{itemize}
\item A close relation between the response function and the clone
  correlation function, which shows analytically 
for the first time how anomalous response requires non trivial clone
correlations. 
\item The derivation of dynamical equations for spherical models that
  avoid the use of the replica method. This method can be generalized
  out of mean-field to obtain a hierarchical system of equations for
  multibody correlation and response functions. 
\item The results of explicit integration of the equation of motion
in spherical spin-glass models, confirming the asymptotic analysis of
the long chain limit and  showing that this limit is achieved in
relatively short chains. 
\end{itemize}
In addition we discussed the derivation of dynamical Ornstein-Zernike
equations suggested by the formalism and we have showed that they have a
structure that generalizes the one of the Mode Coupling equations. These
equations can be closed using schemes borrowed from
equilibrium liquid theory. 
We showed that if the Szamel's closure scheme is applied one recovers the
G\:otze MCT equation. An alternative is the HNC approximation which allows in principle
a quantitative description of aging phenomena in supercooled liquids. 

Within BPD, all the available approximations allowing to describe long
time aging dynamics coherently confirm the original analysis of simple
mean-field spin glass models, in particular effective temperatures
associated to mutual equilibration of slow degrees of freedom
naturally emerge and are interpreted.  We believe that the principle of
quasi-equilibrium configuration space exploration formalized by
Boltzmann pseudodynamics go beyond the approximations and is at the
heart of a description of slow dynamics in terms of effective
temperatures.

\section*{Acknowledgments}
Financial support has been provided by the European Research Council through grant agreement no. 247328 (CriPheRaSy project) and from the Italian Research Ministry through the FIRB Project No. RBFR086NN1. P.U. acknowledges the financial support of the ERC grant NPRGGLASS. He also acknowledges the Department of Physics of the University of Rome La Sapienza and the LPTMS of the University of Paris-Sud 11 where part of this work has been done.

%\bibliography{BIBLIO}

\end{document}